\documentclass[11pt]{article}
\usepackage{moriond,epsfig,here}

\def\Journal#1#2#3#4{{#1} {\bf #2}, #3 (#4)}

\def\NPB{{\em Nucl. Phys.} B}

\def\PRD{{\em Phys. Rev.} D}

\def\be{\begin{equation}}
\def\ee{\end{equation}}
\def\bea{\begin{eqnarray}}
\def\eea{\end{eqnarray}}

\def\D0{\mbox{D$\not\!\!{\rm O}\,$}}
\def\met{\mbox{$\not\!\!{E}_{T}\,$}}

\begin{document}
\vspace*{4cm}
\title{TOP QUARK PRODUCTION CROSS-SECTION AT THE TEVATRON RUN 2}

\author{S.CABRERA,
for the CDF and D\O\ collaborations.}

\address{Department of Physics, Box 90305, Duke University, Durham, NC 27708-0305, USA}

\maketitle\abstracts{
The top quark pair production cross-section ${\sigma}_{t\bar{t}}$ has been measured in $p\bar{p}$ collisions
at center of mass energies of 1.96 TeV using Tevatron Run 2 data.
In the begining of Run 2 both CDF and D\O\ $\sigma_{t\bar{t}}$ measurements in the {\it dilepton} 
channel $t\bar{t}{\rightarrow}WbW\bar{b}{\rightarrow}\bar{\ell}{\nu}_{\ell}b{\ell}^{'}\bar{\nu}_{{\ell}^{'}}\bar{b}$
% $t({\rightarrow}W^{+}b)\bar{t}({\rightarrow}W^{-}\bar{b}){\rightarrow}({\ell}^+{\nu}b),({\ell}^-\bar{\nu}\bar{b})$
and in the {\it lepton plus jets} channel 
$t\bar{t}{\rightarrow}WbW\bar{b}{\rightarrow}q\bar{q}^{'}b{\ell}\bar{\nu}_{\ell}\bar{b}+\bar{\ell}{\nu}_{\ell}bq\bar{q}^{'}\bar{b}$ 
% $t\bar{t}{\rightarrow}({\ell}{\nu}b),({q}^{'}\bar{q}\bar{b})$
agree with the NLO (Next-to-Leading-Order) theoretical predictions.
The presence of a top signal in Tevatron data has been reestablished.}

\section{Introduction}
To date, all direct measurements of top quark production have been performed by the CDF
and D\O\ experiments at the Fermilab Tevatron collider in $p\bar{p}$ collisions.
At the Tevatron top quarks are produced predominatly in pairs through the QCD proceses
$q\bar{q}{\rightarrow}t\bar{t}$ and $gg{\rightarrow}t\bar{t}$. Top quarks can also be produced singly via the electroweak vertex Wtb with about half the cross section, but with final states difficult to extract from background.

In Run 1, at center of mass energies $\sqrt{s}=1.8$ TeV, the top pair production cross-section was expected to be
$5.19^{+0.52}_{-0.68}$ pb at $m_{top}=175~GeV/c^2$ \footnote{ Results BCMN \cite{theobcmn} updated in \cite{theomlm}
taking into account the most recent determinations of systematic uncertainties in the extraction of the PDFs.}
with a $90\%$ ($20\%$) contribution from $q\bar{q}{\rightarrow}t\bar{t}$ ($gg{\rightarrow}t\bar{t}$). 
The precision of the measured cross-sections by the Tevatron from about 100 ${\rm pb}^{-1}$ of data in Run 1
was approximately 25$\%$ \footnote{The ${\sigma}_{t\bar{t}}$ from all channels combined measurement by CDF \cite{cdfxsrun1} and  D\O\ \cite{d0xsrun1} was $6.5^{+1.7}_{-.14}$ pb for $m_{top}=176.1{\pm}6.6~GeV/c^2$ and $5.9\pm1.7$ pb for $m_{top}=172.1{\pm}6.8~GeV/c^2$ respectively.}.
The ratio of cross-sections at 1.96 TeV (Run 1) and 1.8 TeV (Run 2) is 1.295$\pm$0.015, with an expected Run 2 cross-section of 
$6.70^{+0.71}_{-0.88}$ pb with $85\%$ ($15\%$) contribution from $q\bar{q}{\rightarrow}t\bar{t}$ ($gg{\rightarrow}t\bar{t}$) \cite{theomlm}.
In Run 2a, with the increased center of mass energy and the expected
integrated luminosity of 2 ${\rm fb}^{-1}$ we should measure ${\sigma}_{t\bar{t}}$ to better than 
7$\%$ precision and observe single top production for the first time with a $20\%$ precision on the cross-section measurement \cite{mkruse}.

Within the SM the top quark decays almost exclusively into Wb. 
The $t\bar{t}$ {\it dilepton} channel, where both W's decay leptonically to e or $\mu$, has the smallest BR: $5\%$. 
In the so called {\it ``lepton plus jets''} one W decays leptonically and the other hadronically giving a higher BR: ${\sim}30\%$.

The Tevatron has delivered about 170 ${\rm pb}^{-1}$ in Run 2a up until January 2003. The detector upgrades have been extensive.
CDF has  expanded the silicon coverage and installed a new drift chamber. D\O\ has a new inner tracking (silicon and fiber trackers) 
with a new 2T superconducting solenoid. CDF has extended the electron identification to rapidity regions $|{\eta}|>1$ with a new plug calorimeter
and the coverage of the muon systems.

\section{$\sigma_{t\bar{t}}$ measurements in the {\it dilepton} channel}

The background processes that mimic the top {\it dilepton} signature are Drell-Yan ($Z^{*}/{\gamma}{\rightarrow}e^+e^-,{\mu}{\mu}$), $Z{\rightarrow}{\tau}{\tau}$, WW/WZ and processes with a real lepton and a jet or a track {\it faking} a second lepton.

Dilepton selection starts with 2 high-$P_t$ ($P_t>20~GeV/c$) $e$ or $\mu$ oppositely charged. CDF requires both leptons to be well isolated 
from nearby calorimeter activity greatly reducing the {\it fake lepton}, W$b\bar{b}$ and $b\bar{b}$ backgrounds.
The dilepton invariant mass, $M_{ee}$ or $M_{{\mu}{\mu}}$, is required to be outside the interval $76-106~GeV/c^2$ to reject $Z{\rightarrow}{\ell}^{+}{\ell}^{-}X$ events. D\O\ discriminates $t\bar{t}$ from Z's in this interval by demanding larger \met\footnote{ \met is the missing energy transverse to the beam direction.} than in the region outside this interval.

\begin{figure}[hbt]
\label{fig:tt}
\begin{center}
\begin{tabular} {cc}
    \mbox{\epsfysize=6.2cm\epsfxsize=6.5cm
\epsffile{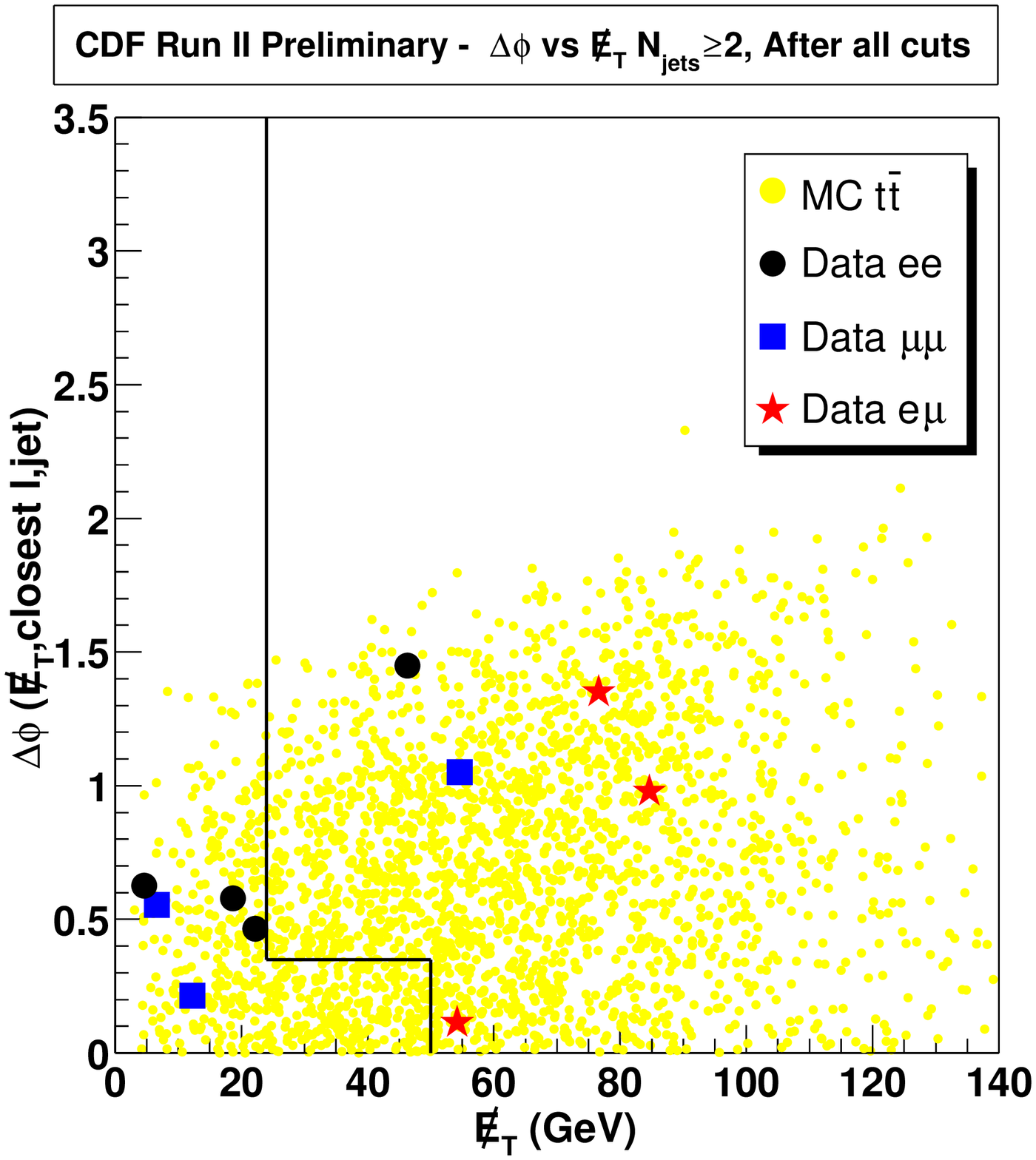}} &
    \mbox{\epsfysize=6.2cm\epsfxsize=6.5cm
\epsffile{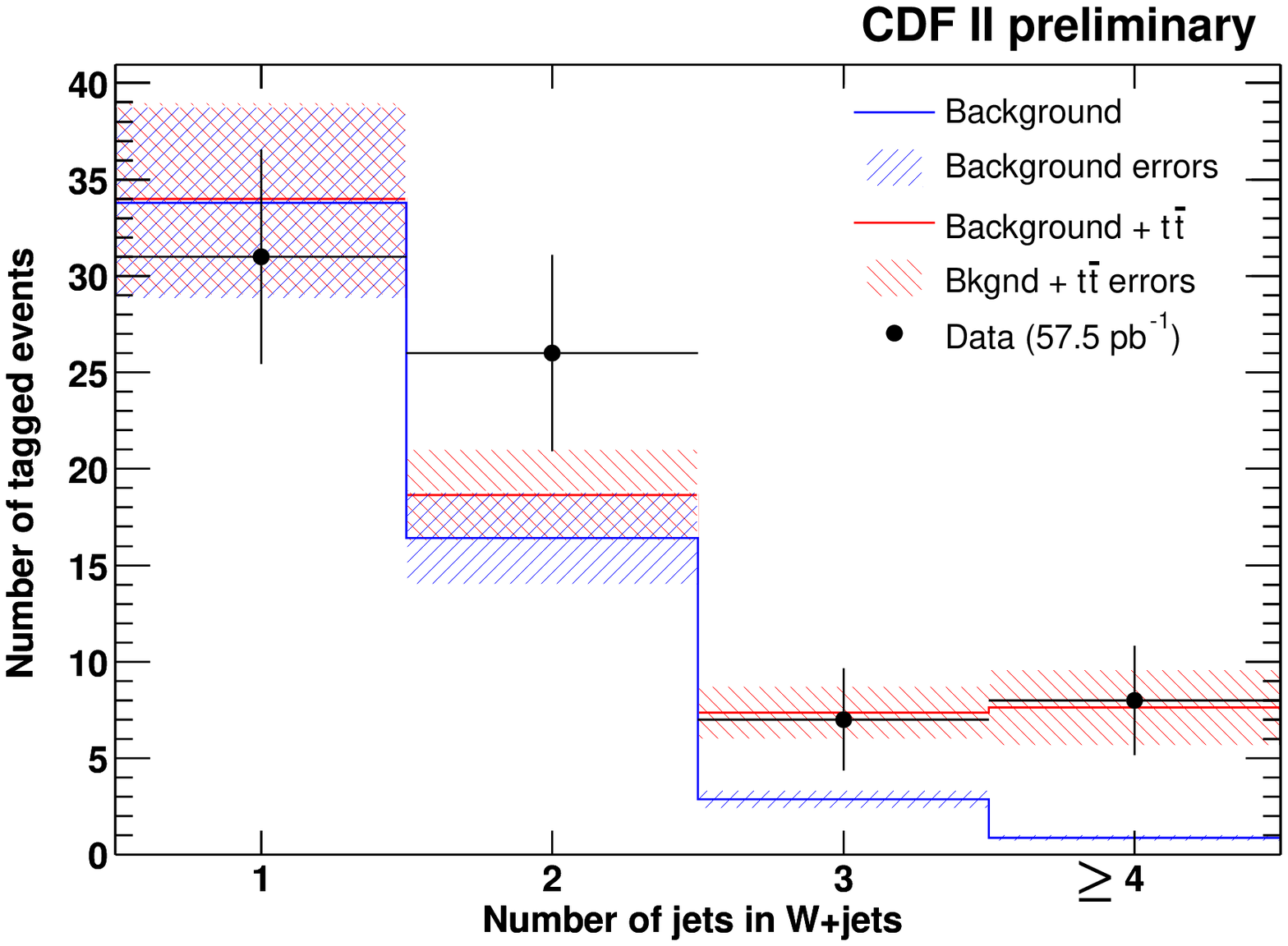}}
\\
\end{tabular}
\end{center}
\vskip -0.5cm
\caption{On the left side, the 5 CDF $t\bar{t}$ dilepton candidates found in 72 $pb^{-1}$ in the 
plane ${\Delta}{\phi}(\met,nearest~{\ell}~{\rm or}~j)$ versus \met in comparison with MC Herwig $t\bar{t}$. On the right side,
number of events in the $W+{\it jets}$ sample with at least one b-tag: the 3 and ${\ge}4$ jet bins are used to extract $\sigma_{t\bar{t}}$.}
\end{figure}

A large \met is required as a signature of the two W decay neutrinos. All backgrounds with real \met contribution due to the presence of neutrinos are reduced. In addition CDF requires $|{\met}|>50~GeV$ if ${\Delta}{\phi}(\met,{\ell}~{\rm or}~j)<20^{0}$ to eliminate instrumental contributions to the \met due to mismeasured energies of lepton or jets \footnote{${\Delta}{\phi}(\met,{\ell}~{\rm or}~j)$ is  the azimutal separation between the vector \met and the nearest lepton or jet.} (see Figure
\ref{fig:tt}). 

\vskip -0.15cm

%%%%%%%%%%%%%%%%%%%%%%%%%%
%%%%%%%%%%%%%%%%%%%%%%%%%%

\vskip -0.15cm

Two high energy jets are demanded as expected by the fragmentation of the top decay b quarks. The backgrounds with softer jets originating from QCD 
radiation are reduced. Finally, to enhance the signal-to-background ratio, large $H_T$ is required 
\footnote{$H_T$ is the scalar sum of the transverse energy of the leptons, jets and neutrinos in the event.}. 
The results from CDF and D\O\ are summarized in Tables \ref{tab:cdf_dil} and \ref{tab:d0_dil}.

%TABLAS

\begin{table}[hbt]
\caption{Run 2 CDF results in the $t\bar{t}$ dilepton channel for a data sample of 72 $pb^{-1}$ \label{tab:cdf_dil}}
\vspace{0.4cm}
\begin{center}
\begin{tabular}{|l|c|c|c|c|}
\hline
Source & ee & ${\mu}{\mu}$ & $e{\mu}$ & ${\ell}{\ell}$ \\ \hline
All Backgrounds & 0.103$\pm$0.056 &  0.093$\pm$0.054 &  0.100$\pm$0.037 & 0.30$\pm$0.12 \\ \hline 
Expected  $t\bar{t}{\rightarrow}{\ell}{\nu}_{\ell}b{\ell}^{'}\bar{\nu}_{{\ell}^{'}}\bar{b}$
& 0.47$\pm$0.05 &  0.59$\pm$0.07 &  1.44$\pm$0.16 & 2.5$\pm$0.3 \\ \hline
Data & 1 & 1 & 3 & 5 \\ \hline
\end{tabular}
\end{center}
\end{table}

\begin{table}[ht]
\caption{Run 2 D\O\ results in the $t\bar{t}$ dilepton channel \label{tab:d0_dil}}
\vspace{0.4cm}
\begin{center}
\begin{tabular}{|l|c|c|c|c|}
\hline
Source & ee & ${\mu}{\mu}$ & $e{\mu}$ \\ \hline
${\cal L}~{\rm pb}^{-1}$ & 48.2 & 42 & 33 \\ \hline
All Backgrounds   
& 1.00$\pm$0.48 &  0.6$\pm$0.30 &  0.07$\pm$0.01 \\ \hline
Expected  $t\bar{t}{\rightarrow}{\ell}{\nu}_{\ell}b{\ell}^{'}\bar{\nu}_{{\ell}^{'}}\bar{b}$
& 0.25$\pm$0.02 &  0.3$\pm$0.04 &  0.50$\pm$0.01 \\ \hline
Data & 4 & 2 & 1 \\ \hline
\end{tabular}
\end{center}
\end{table}

\section{$\sigma_{t\bar{t}}$ measurements in the {\it lepton+jets} channel}

The CDF event selection required one $e$ or $\mu$ with $P_t>20~GeV/c$, $\met>20~GeV$ and at least 3
high ${\rm E}_{t}$ jets. Cosmic rays, electron conversions, Drell-Yan and $t\bar{t}$ dilepton events are removed. 

To increase the signal-to-background ratio, CDF uses the Silicon Vertex Detector to identify the b-quark displaced vertices. A jet
is b-tagged if it contains a secondary vertex with at least two charged tracks and $\frac{L_{xy}}{{\sigma}_{xy}}>3$ 
\footnote{$L_{xy}$ is the distance in the transverse plane to the beam direction between the secondary vertex and the primary vertex. ${\sigma}_{xy}$ is 
the resolution in the determination of $L_{xy}$.}. The efficiency for identifying at least one of the b quarks from $t\bar{t}$ decays is about 45$\%$, 
which is measured using $t\bar{t}$ MC and corrected with a data to MC scale factor. 
The mistags from light quarks and gluon jets are evaluated using the negative rate of $L_{xy}$ extracted from inclusive jet data
and applied to W+{\it jets} data. The W/Z+heavy flavour: g${\rightarrow}b\bar{b},c\bar{c}$ background is evaluated from W+{\it jets} data, the b tag rate and the Run 1 flavour composition in W+{\it jets} events. 
The non-W background is evaluated from W+{\it jets} data assuming it is flat in the plane of lepton calorimeter isolation versus \met, and extrapolated 
from the low isolation and small \met (non-W) region to the high isolation and large \met (W dominated) region.   
Small contributions from diboson WW/WZ, Drell-Yan and single top production are evaluated from MC (see results in Table \ref{tab:cdf_ljets}). 

\begin{table}[htb]
\caption{Run 2 CDF results in the $t\bar{t}$ {\it lepton plus jets} channel with displaced vertex tagging \label{tab:cdf_ljets}}
\vspace{0.4cm}
\begin{center}
\begin{tabular}{|l|c|c|c|c|}
\hline
Source & W+1{\it jet} & W+2{\it jets} & W+3{\it jets}& W${\ge}4${\it jets} \\ \hline

Background   

& 33.8$\pm$5.0  & 16.4$\pm$2.4 & 2.88$\pm$0.05 & 0.87$\pm$0.2 \\ \hline

SM Background plus $t\bar{t}$

& 34.0$\pm$5.0 &  18.65$\pm$2.4 &  7.35$\pm$1.4 & 7.62$\pm$2.0 \\ \hline

Data before tagging & 4913 & 768 & 99 & 26 \\ \hline
Data (${\ge}1$b-tag) & 31 & 26 & 7 & 8 \\ \hline

\end{tabular}
\end{center}
\end{table}

The D\O\ {\it topological} analysis does not use b-tagging. First, a data sample enriched with W events
is preselected by demanding a loose e or $\mu$ with $P_t>20~GeV/c$, $\met>20~GeV$ and a Soft Muon Tag veto. Then, the QCD background
is evaluated from data for each jet multiplicity. In the $e$-channel this background is due to $\pi^{0}$ and $\gamma$ QCD compton in jets
{\it faking} $e$'s and in the ${\mu}$-channel is due to real ${\mu}$'s from heavy flavour decays. The $W+{\ge}4~{\it jets}$ background 
is estimated using the Berends scaling law. Finally, the topological cuts are applied to further reduce background: at least 4 jets, 
and large values of $H_t$ and $\cal A$ \footnote{The Aplanarity $\cal A$ measures the relative activity perpendicular to the plane of maximum 
activity.} (see results in Table \ref{tab:d0_kin_ljets}). 

The D\O\ {\it Soft Muon Tag} analysis has same preselection as the {\it topological} analysis. The topological requirements on $H_t$ and $\cal A$ are milder and at least 3 high-$E_{t}$ jets are required. Background is reduced by demanding one low momentum $\mu$ in a jet coming from the semileptonic b decay (see results in Table \ref{tab:d0_smt_ljets}).

\vskip -0.2cm

\begin{table}[htb]
\caption{Run 2 D\O\ results in the $t\bar{t}$ {\it lepton plus jets} {\it topologic} analysis \label{tab:d0_kin_ljets}}
\vspace{0.4cm}
\begin{center}
\begin{tabular}{|l|c|c|c|c|c|c|}
\hline
        & ${\rm N}_W$ & ${\rm N}_{\rm QCD}$ & All BG & Exp Signal & ${\rm N}_{\rm obs}$ & ${\cal L}(pb^{-1})$ \\ \hline
e+jets       & 1.3$\pm$0.5 & 1.4$\pm$0.4  & 2.7$\pm$0.6 & 1.8 & 4 & 49.5 \\ \hline
${\mu}$+jets & 2.1$\pm$0.9 & 0.6$\pm$0.4  & 2.7$\pm$1.1 & 2.4 & 4 & 40 \\ \hline

\end{tabular}
\end{center}
\end{table}

\vskip -0.3cm

\begin{table}[htb]
\caption{Run 2 D\O\ results in the $t\bar{t}$ {\it lepton plus jets} {\it Soft Muon Tag} analysis \label{tab:d0_smt_ljets}}
\vspace{0.4cm}
\begin{center}
\begin{tabular}{|l|c|c|c|c|}
\hline
             & All BG      & Exp Signal   & ${\rm N}_{\rm obs}$ & ${\cal L}(pb^{-1})$ \\ \hline
e+jets       & 0.2$\pm$0.1 & 0.5          & 2                   & 49.5 \\ \hline
${\mu}$+jets & 0.7$\pm$0.4 & 0.8          & 0                   & 40 \\ \hline

\end{tabular}
\end{center}
\end{table}

\section{Summary and conclusions}
The $t\bar{t}$ production cross-section in $p\bar{p}$ collisions at $\sqrt{s}=1.96$ TeV has been determined from 
the number of observed top candidates in a given channel, the estimated background, the integrated luminosity 
and the $t\bar{t}$ acceptance for a top mass 175 $GeV/c^2$ \footnote{For the latest results on top mass see \cite{rzitoun}.}: $\sigma_{t\bar{t}}=\frac{N_{obs}-N_{bkg}}{A\cdot{\int {\cal L}}}$. All results are in agreement with the NLO prediction: $6.70^{+0.71}_{-0.88}$ pb. 
Attributing the excess of events over the expected backgrounds to $t\bar{t}$ production in the decay channels considered, we obtain the following first Run 2 results:

\begin{itemize}
\item CDF {\it dilepton channels}: $\sigma_{t\bar{t}}=13.2\pm5.9(stat)\pm1.5(sys)\pm0.8(lum)$ pb.
\item CDF {\it lepton plus jets channels}: $\sigma_{t\bar{t}}=5.3\pm1.9(stat)\pm0.8(sys)\pm0.3(lum)$ pb.
\item D\O\ {\it dilepton channels}: $\sigma_{t\bar{t}}=29.9^{+21.0}_{-15.7}(stat)^{+14.1}_{-6.1}(sys)\pm3.0(lum)$ pb.
\item D\O\ {\it lepton plus jets channels}: $\sigma_{t\bar{t}}=5.8^{+4.3}_{-3.4}(stat)^{+4.1}_{-2.6}(sys)\pm0.6(lum)$ pb.
\item D\O\ {\it all combined channels}: $\sigma_{t\bar{t}}=8.5^{+4.5}_{-3.6}(stat)^{+6.3}_{-3.5}(sys)\pm0.8(lum)$ pb.
\end{itemize}

\section*{Acknowledgments}
Thanks to all the participants in the CDF and D\O\ top physics groups.
Their extraordinary amount of effort made the Winter 2003 measurements reported
in this proceedings a reality. Thanks to P. Azzi-Bachetta, J. Konigsberg, E. Barberis and C. Gerber to 
motivate and coordinate all this effort.

\end{document}